\documentclass[aps,prl,twocolumn,showpacs,superscriptaddress]{revtex4-1}

\usepackage{graphics}
\usepackage{amsmath}
\usepackage{amsfonts}
\usepackage{bm}

\usepackage{graphicx,mdwlist,color,amsmath,wrapfig,revsymb}

\begin{document}

\title{Polarization-dependent photocurrents in polar stacks of van der Waals solids}

\author{Y. B. Lyanda-Geller}
\affiliation{Department of Physics and Astronomy, Purdue University, West Lafayette, Indiana 47907 USA}
\affiliation{Birck Nanotechnology Center, Purdue University, West Lafayette, Indiana 47907 USA}

\author {Songci Li}

\affiliation{Department of Physics, University of Washington, Seattle, WA 98195 USA}

\author{A. V. Andreev}
\affiliation{Department of Physics, University of Washington, Seattle, WA 98195  USA}

\date{September 7, 2015}
\pacs{73.63.-b, 78.67.- n, 73.50.Pz, 72.15.Gd}

\begin{abstract}

Monolayers of semiconducting van der Waals solids, such as transition metal dichalcogenides (TMDs),
acquire significant electric polarization normal to the layers when placed  on a substrate or
in a heterogeneous stack. This causes linear coupling of electrons to electric fields normal to the layers. Irradiation at oblique incidence at frequencies above the gap
causes  interband transitions due to coupling to both normal and in-plane ac electric fields. The interference between
the two processes leads to sizable  in-plane photocurrents and valley currents.  The direction and magnitude of currents is controlled by light polarization and is determined by its helical or nonhelical components. 
The helicity-dependent ballistic current arises due to asymmetric photogeneration.
The non-helical current has a ballistic contribution (dominant in sufficiently clean samples)
 caused by asymmetric scattering of photoexcited carriers, and a side-jump contribution. Magneto-induced photocurrent is due to
the Lorentz force or due to intrinsic magnetic moment related to Berry curvature.

\end{abstract}

\maketitle

{\it Introduction.}  Since the discovery of graphene~\cite{GeimNobel} a whole class of novel two
dimensional (2D) materials, called van der Waals solids (vdWs),
has been identified~\cite{GeimNovoselov_vdW,GeimGrigorieva}.
In these materials 2D monolayers with strong in-plane bonding are coupled by weak van der Waals interactions.
Few-monolayer thick structures of vdW materials have electronic and optical properties that can
differ drastically from those of the bulk phases~\cite{heinz2010}.
vdW materials exhibit phenomena associated with valley degrees of freedom,
such as valley Hall currents \cite{niu} and valley-selective carrier
photoexcitation by circularly polarized  light \cite{Xu},
related to topological properties of the bands, such as Berry curvature
and valley-dependent magnetic moment \cite{heinz2014}.
Stacking of monolayers of different vdW solids enables fabrication
of novel artificial structures with interesting electronic properties~\cite{GeimNovoselov_vdW,GeimGrigorieva}.

In their natural form most vdW materials are nonpolar. When monolayers of different vdW materials are stacked
in a heterostructure or placed on a substrate, an electric dipole moment perpendicular to the layers arises. This allows for photogalvanic effects (PGE): electric currents due to illumination by light in the absence of external electric field. In particular, a photocurrent arises between the top and bottom contacts of a heterojunction fabricated from monolayers of different vdW solids~\cite{vanderZande}. This current does not depend on the polarization of light, is caused by spatial separation of photoexcited electrons and holes in a junction, and belongs to a class of effects in which the direction of the photocurrent is governed by spatial inhomogeneities of the sample or its illumination. Another class of effects,  in which the direction of photocurrent or photovoltage is determined by the polarization of light~\cite{Belinicher1980,Ivchenko2005},  occurs even in  uniformly illuminated spatially homogeneous solids.
Recently, polarization-sensitive  photocurrents were observed~\cite{Hwang} when the 2D conduction layer formed  at the interface of a  WSe$_2$ stack and the substrate was irradiated at frequencies  below the band gap.

Here we show that coupling of radiation to the electric dipole moment in stacks of undoped semiconducting vdW solids leads to sizable polarization-dependent photocurrents for frequencies above the band gap. Quantum interference between this coupling and electron coupling to the in-plane electric field component of the radiation results in carrier photogeneration in the conduction and valence bands that leads to a valley and net  currents.

\begin{figure}
	\includegraphics[height=2in,width=3in]{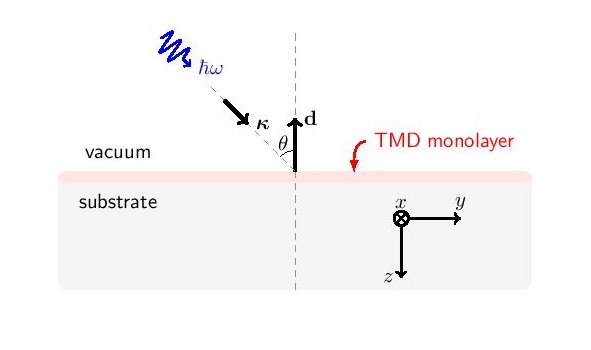}
	\caption{\label{fig:setup} (color online) Schematic representation of the system. Irradiation of a semiconducting  polar TMD monolayer by helical light at oblique incidence, $\theta\neq 0 $, generates a helicity-dependent net photocurrent perpendicular to the plane of incidence $yz$. For linear polarization, a net current is generated in the plane of incidence. }	
\end{figure}

For 2D structures with reflection symmetry broken by the dipole moment,  polarization-dependent PGE currents, to linear order in light intensity, can be expressed in terms of a polar vector $\mathbf{d}=(0,0,d_z)$ perpendicular to the layers by the phenomenological relation
\begin{equation}\label{eq:j_symmetry}
\mathbf{j}=\xi \,  \mathbf{d} \times i  \left[  \mathbf{E}\times \mathbf{ E}^* \right]
+ \zeta \left[\mathbf{E}^*(\mathbf{d} \cdot \mathbf{E})+\mathbf{E}\,(\mathbf{d} \cdot  \mathbf{ E}^*)\right]
.
\end{equation}
Here $\mathbf{E}$ is the complex electric field amplitude of a monochromatic light, $\mathbf{E}(t)=\Re (\mathbf{E} e^{-i \omega t})$, and  the (real) phenomenological parameters  $\xi$ and $\zeta$ describe, respectively the \textit{circular} and \textit{linear} PGE. In Eq.~(\ref{eq:j_symmetry}) the electric field of the radiation is assumed spatially uniform and the photon momentum is neglected.  The in-plane photocurrent  arises when the sample is illuminated  at oblique incidence, as shown in Fig.~\ref{fig:setup}, and its direction and magnitude are determined by the polarization of light.

Determination of the physical mechanism of the photocurrent and evaluation of the phenomenological parameters  $\mathbf{d}$, $\xi$ and $\zeta$ in Eq.~(\ref{eq:j_symmetry}) requires a microscopic theory.  The photocurrents arise due to:  i) asymmetric photoelectron generation (with different generation rate for opposite electron momenta) \cite{Ivchenko1978,Belinicher1978,ylg1984}, and ii)  asymmetric kinetics  (when light-induced symmetric momentum distribution leads to the current due to asymmetric scattering, due to side jumps,   spin relaxation, or  evolution in magnetic field), \cite{Belinicher1982,Ivchenko,Dyakonov1982,Lyanda-Geller1987,LyandaGeller1989,Ivchenko1989,Ivchenko1990}. These mechanisms describe well \ experiments detecting polarization-dependent currents in bulk semiconductors, such as Te \cite{Asnin1978} and GaAs \cite{Andrianov1984,Bakun1984,Andrianov1992}, and photocurrents in III-V type heterostructures \cite{Ganichev2001}.
  Both asymmetric photogeneration and kinetics play an important role in the discussion below.

When the mean free path of the photoexcited carriers exceeds their de Broglie wavelength, the photocurrent (\ref{eq:j_symmetry})  can be expressed in terms of  the electron distribution function $f_l(\mathbf{p}, \mathbf{r})$  as~\cite{Ivchenko,Vignale,Aleiner}
\begin{equation}
\mathbf{j}=e\sum_{\mathbf{p},l} \left[ \mathbf{v}_l(\mathbf{p}) +\delta\mathbf{v}_l(\mathbf{p})\right] f_{l}(\mathbf{p}).
\label{current}
\end{equation}
The first term here with the group velocity in a band $l$ $\mathbf{v}_l(\mathbf{p})= \partial \epsilon_l(\mathbf{p})/\partial\mathbf{p}$ describes the ballistic current. The second term is the side jump (shift) current due to the displacement $\mathbf{R}_{l',l}(\mathbf{p}',\mathbf{p})$ of the center of mass of the wave packet 
  during a transition from state $l,\mathbf{p}$ to state $l', \mathbf{p}'$ (as a result of scattering~\cite{Luttinger,Berger} or photoabsorption~\cite{Ivchenko,Lyanda-Geller1987}).
The correction to the velocity is expressed in terms of the transition probability $W_{l',l}(\mathbf{p}',\mathbf{p})$ as
\begin{equation}
\label{eq:delta_v}
\delta \mathbf{v}_{l}(\mathbf{p})=\sum_{l',\mathbf{p}'}W_{l',l}(\mathbf{p}',\mathbf{p}) \mathbf{R}_{l',l}(\mathbf{p}',\mathbf{p}).
\end{equation}
The magnitude of the side jump is expressed in terms of the phase of the transition  matrix element $T_{l',l}(\mathbf{p}',\mathbf{p})$ as
\begin{equation}
\label{eq:side_jump}
\mathbf{R}_{l',l}(\mathbf{p}',\mathbf{p})= \mathbf{\Omega}_{l'} (\mathbf{p}') - \mathbf{\Omega}_{l} (\mathbf{p})-\left( \partial_\mathbf{p} + \partial_{\mathbf{p}'} \right) \Im \ln T_{l',l}(\mathbf{p}',\mathbf{p}) ,
\end{equation}
where $\mathbf{\Omega}_{l} (\mathbf{p})$ is the Berry connection in band $l$.  Gauge invariance of
 (\ref{eq:side_jump}) is obvious: When $\mathbf{\Omega}_{l} (\mathbf{p}) \to \mathbf{\Omega}_{l} (\mathbf{p}) - \partial_{\mathbf{p}}\chi_l(\mathbf{p})$,   $T_{l',l}(\mathbf{p}',\mathbf{p})  \to  T_{l',l}(\mathbf{p}',\mathbf{p})\, e^{ i \chi_l(\mathbf{p}) - i  \chi_{l'}(\mathbf{p}')}$.

In a spatially uniform steady state, and in the absence of static external  electric and magnetic fields, the
nonequilibrium part $\delta f_l (\mathbf{p})$ of the electron distribution function is determined by the balance between
the photogeneration due to direct interband transitions $J_l(\mathbf{p})$ and relaxation and recombination of photoexcited carriers,
\begin{equation}
\label{eq:collision_balance}
\sum_{l',\mathbf{p}'}w_{l',l}(\mathbf{p}',\mathbf{p})
\left[\delta f_{l'}(\mathbf{p}')-\delta f_{l}(\mathbf{p})\right]
+ J_l(\mathbf{p})=0,
\end{equation}
where $w_{l',l}(\mathbf{p}',\mathbf{p})$ is the probability of momentum relaxation \cite{mixed}. Below we apply Eqs.~(\ref{current}) and (\ref{eq:collision_balance}) to the study of polarization-dependent currents  (\ref{eq:j_symmetry}) in polar stacks of semiconducting TMDs,  such as MoS$_2$ and WSe$_2$.

\begin{figure}
	\includegraphics[height=2in,width=3in]{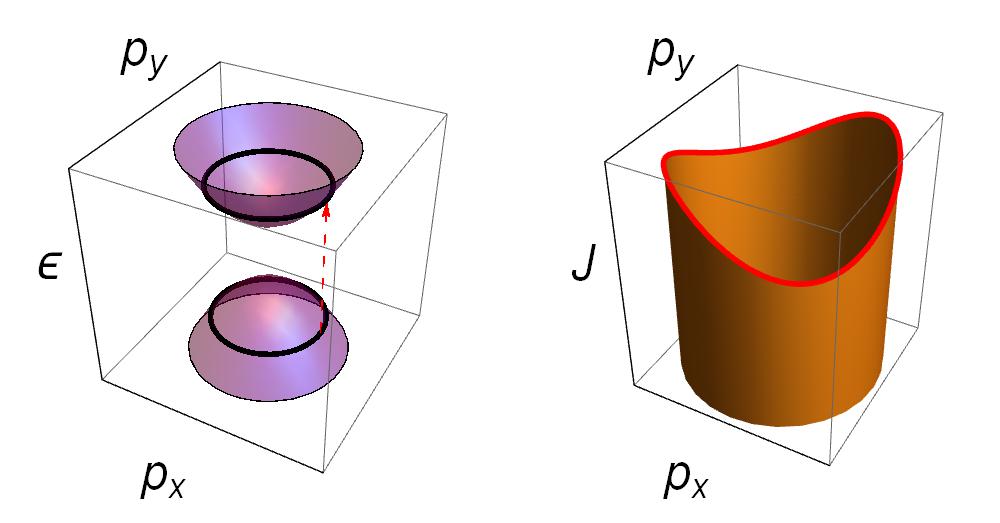}
	\caption{\label{fig:photogeneration}(Color online). Left: Direct transitions (red arrow) between electron states in the valence and conduction bands shown for one of the valleys by black circles. Right: Asymmetric photogeneration rate. }	
\end{figure}

{\it Asymmetric photogeneration.}  Semiconducting TMD at low number of monolayers are direct band semiconductors with strong coupling to light and sizable charge carrier mobility \cite{heinz2010,heinz2014,Kis}. The inter-layer tunneling is weak and we neglect it.
Since in this approximation the total in-plane photocurrent is the sum of contributions of individual layers, we consider the photocurrent in a single layer of TMD either placed on a substrate or in a polar stack.
We assume that photon energy is not too far from the absorption threshold. In this case only electrons with
momenta near the $K$  and $K^{\prime}$ points of the hexagonal Brillouin zone
absorb light and produce photocurrent, see left panel in Fig.~\ref{fig:photogeneration}. The effective two-band Hamiltonian for such low energy electrons is \cite{niu,Xu,Fal'ko}
\begin{equation}
{\cal H}= v({\mathbf \tau}_z{\mathbf \sigma}_x p_x+{\mathbf \sigma}_y p_y) +\Delta {\mathbf \sigma}_z,
\label{Hamiltonian2D}
\end{equation}
where the momentum $\mathbf{p}$ is measured from the $K$ or $K'$ point,  $v$ has dimensions of velocity, and $\Delta$ is half the bandgap between the spin-nondegenerate conduction and valence bands. The  Pauli matrices ${\mathbf\sigma}_i$ act on the band pseudospin, and ${\mathbf \tau}_z$  acts on the valley pseudospin.

At normal incidence of the radiation, electrons couple to the in-plane component of the ac electric field. The corresponding coupling Hamiltonian  is obtained from Eq.~(\ref{Hamiltonian2D}) by the usual substitution $\mathbf{p}\rightarrow \mathbf{p}-e\mathbf{A}/c$, where $\mathbf{A}$ is the vector potential and $e$ is the electron charge. This results in valley-selective transitions for circularly-polarized light \cite{Xu}; Application of an in-plane dc electric field results in valley current \cite{niu}. At oblique incidence, electrons in a polar stack also couple to the normal component of the ac electric field, $E_z(t)$.
The full coupling of electrons to the (uniform) ac electric field is given by
\begin{equation}
V=  -\frac{e v }{c} \left({\mathbf \tau}_z{\mathbf \sigma}_x A_x+{\mathbf \sigma}_y A_y\right) + \,\frac{1}{c} d_z \dot A_z{\mathbf \sigma}_z,
\label{light}
\end{equation}
where the electric field enters through the time derivative of the vector potential, $\mathbf{E}=-\dot{\mathbf{A}}/c$, and $d_z$ is the difference between the dipole moments of electron states in the conduction and valence bands, which arises as follows. If $E^b_z$ is a  built-in electric field  in a polar TMD stack, the total $z-$component of the electric field is $E^t_z=E^{b}_z+E_z(t)$.
This electric field couples orbitals even in $z$, that form the conduction and valence bands described by (\ref{Hamiltonian2D}),
  to odd in $z$ higher and lower band states with energies $\epsilon_s$, with $s$ labeling odd bands. Then the energies of the bottom of the conduction band and the top of the valence band $\epsilon^0_{c(v)}$ change: $\delta \epsilon_{c(v)}=\sum_s |(eE^t_z z)_{c(v)s}|^2/(\epsilon^0_{c(v)}-\epsilon_s)$. Thus a coupling of charge carriers to light linear in electric field $E_z(t)$ arises, and the dipole moment difference $d_z= e^2\sum_s [(E_z^{b}z)_{cs}z_{sc}/(\epsilon^0_{c}-\epsilon_s)-(E_z^{b}z)_{vs}z_{sv}/(\epsilon^0_{v}-\epsilon_s)]$.
This coupling plays a crucial role in generation of polarization-dependent photocurrent in vdW materials. The value of  $d_z$ can be  estimated from the measured~\cite{vanderZande} dependence of the band gap on the applied external electric field perpendicular to the layers, $d_z = - d \Delta/dE_z $.

Optical transitions between the valence (-) and conduction (+) band in the $K$-valley are described by the matrix elements $V^K_{-+}(\mathbf{p})={\Psi^K_+(\mathbf{p})}^\dagger V\Psi^K_-(\mathbf{p})$, where the wavefunctions $\Psi^K_{\pm}(\mathbf{p})$ corresponding to energies $\pm\epsilon=\pm \sqrt{(vp)^2+\Delta^2}$ are
\begin{equation}
\left(\Psi^K_\pm(\mathbf{p})\right)^T=\left(
\pm v p_-/\sqrt{2(\epsilon\mp \Delta)\epsilon}, \, \sqrt{\epsilon \mp\Delta/2\epsilon}
\right).
\label{pm}
\end{equation}
Here $p_\pm = p_x\pm i p_y$, and the superscript $T$ indicates a matrix transposition.
For the $K^{\prime}$-valley, the wavefunctions are obtained by replacing $p_-$ in Eq.~(\ref{pm}) with $-p_+$.
The rate of direct optical transitions, see Fig.~\ref{fig:photogeneration},  in the$K$ $(j=1)$ or  $K^{\prime}$  $(j=2)$  valley, assuming fully occupied valence band and empty conduction band, can be  determined using the Fermi golden rule,
$J_{+,j}=-J_{-,j}=J_j=\frac{2\pi}{\hbar}|V^j_{-,+}|^2\delta(\hbar\omega-2\epsilon)$, and is given by
\begin{eqnarray}
J_j (\mathbf{p})&&=\frac{2\pi}{\hbar}\left(\frac{ e |\mathbf{E}|v}{\omega}\right)^2 Z(p) \delta (\hbar\omega - 2\epsilon)\times\nonumber \\
&&\left[ \frac{1-|e_z|^2}{2}\frac{\Delta^2 + \epsilon^2}{\epsilon^2} -(-1)^{j}\kappa_z \frac{\Delta}{\epsilon}+ |e_z|^2\frac{(\omega d_z p)^2}{e^2\epsilon^2}\right.\nonumber \\
&&-\frac{v^2}{2\epsilon^2}\left[(|e_x|^2-|e_y|^2)(p_x^2-p_y^2)+ 2S_{xy}p_xp_y\right]+\nonumber \\
&&\left.\frac{\omega }{e\epsilon} \, \mathbf{p} \cdot\left( \frac{\Delta}{\epsilon} \left[ \boldsymbol{\kappa}\times\mathbf{d} \right]
+(-1)^j  [\hat{\boldsymbol{z}}\times \hat{S}\mathbf{d}]
\right)\right].
\label{eq:photogeneration}
\end{eqnarray}
Here $\hat{\boldsymbol{z}}$ is the $z$-axis unit vector, $\mathbf{e}=\mathbf{E}/|\mathbf{E}|$ is the light polarization vector, and the pseudovector $\boldsymbol{\kappa}=i\mathbf{e}\times\mathbf{e}^*$ and the tensor  $\hat S$, $S_{ij}=e_ie_j^*+e_i^*e_j$, characterize, respectively, the helical and the non-helical components of light polarization, with $(\hat{S}\mathbf{d})^T=(S_{xz},S_{yz},S_{zz})d_z$.  The Sommerfeld factor~\cite{Landau3} $Z(p)$   accounts for the Coulomb interaction between the photogenerated electron and hole.  In the 2D case for a quadratic energy dispersion,
 $Z(p)= 2\left[1+ \exp\left(- 2\pi \hbar/p a_B\right)\right]^{-1}$~\cite{Shinada1966}, where $a_B=\hbar^2 \varepsilon/\mu e^2$ is the exciton Bohr radius, $\varepsilon$ is the dielectric constant,  $p$ is the electron or hole momentum, and $\mu$ is the reduced effective mass. In our model, $p=\sqrt{\epsilon^2 - \Delta^2}/v$, $\mu=\Delta/v^2$.

The momentum dependence of the photogeneration in Eq.~(\ref{eq:photogeneration}) is illustrated in the right panel of Fig.~\ref{fig:photogeneration}.
The asymmetry of photogeneration responsible for the in-plane photocurrent arises  from the interference between coupling of electrons to the in-plane electric field of light  and the linear Stark coupling to the normal field $E_z$ caused by the dipole $\mathbf{d}$. It is described by the two terms linear in the electron momentum in last line of Eq.~(\ref{eq:photogeneration}).

The different angular harmonics of the nonequilibrium distribution function relax independently. Therefore for the photocurrent it is sufficient to consider the first angular harmonic $\delta f^{(1)}_l (\mathbf{p})$ of the nonequilibrium distribution function,
$
\delta f^{(1)}_l (\mathbf{p}) =  \left(\mathbf{A}_l  + \tau_z \mathbf{B}_l \right) \cdot \hat{\mathbf{p}}$, where $\mathbf{A}_l$ and  $\mathbf{B}_l$ characterize valley-even and odd asymmetry of momentum distribution, respectively,
and $\hat{\mathbf{p}}=\mathbf{p}/|\mathbf{p}|$ is a unit vector along the electron momentum. The relevant scattering probability in Eq.~(\ref{eq:collision_balance}) is given by
\begin{equation}
\label{eq:w_skew}
w_{l',l} (\mathbf{p}',\mathbf{p})\to\delta_{l,l'}\left[ \frac{1}{\tau_l} \,  \hat{\mathbf{p}}  \cdot \hat{\mathbf{p}}' + \tau_z \frac{1}{\tau_l^{sk}} \hat{\boldsymbol{z}} \cdot \hat{\mathbf{p}} \times \hat{\mathbf{p}} ' \right],
\end{equation}
where $1/\tau_l$ and $1/\tau_l^{sk}$ are respectively the transport and skew momentum relaxation rates in band $l$.

\textit{Ballistic photocurrent.} The first term in Eq.~(\ref{current}) for the photocurrent describes charge transfer during ballistic motion of electrons, and  is characterized by the  asymmetric in momentum part of the distribution function. 
The latter is caused by asymmetric photogeneration or subsequent asymmetric  scattering.
The ballistic circular PGE arises directly due to the valley-even asymmetric photogeneration (first term in the last line of Eq.~(\ref{eq:photogeneration})). We find that the dominant ballistic linear PGE requires a conversion, via skew scattering, of the valley-odd photogeneration (last term in Eq.~(\ref{eq:photogeneration})) into a valley-even asymmetric  momentum distribution. Skew scattering arises only in the second Born approximation. As a result, 
although both transport and skew scattering rates are proportional to the impurity concentration, 
the skew scattering rate is smaller in the parameter $\tau_l/\tau_l^{sk}\sim \delta_l \ll 1$, where $\delta_l$ is a phase shift of electron scattering off impurities in a band $l$. The ballistic contribution to linear and circular PGE coefficients $\xi$ and $\zeta$ in Eq.~(\ref{eq:j_symmetry}) are given by
\begin{subequations}
\label{eq:xi_zeta_bal}
\begin{eqnarray}
\xi^{bal} &=& \left(\frac{e}{\hbar }\right)^2\frac{Z(p_\omega)\left[(\hbar\omega)^2-(2\Delta)^2\right]\Delta }{(\hbar \omega)^3}\, (\tau_c+\tau_v), \label{eq:xi_bal}\\
\zeta^{bal}&=&
\xi^{bal}\frac{\hbar \omega}{\Delta \left(\tau_c+\tau_v \right)}
 \left(\frac{\tau_c^2}{\tau^{sk}_{c}}+\frac{\tau^2_v}{\tau^{sk}_{v}}\right). \label{eq:zeta_bal}
\end{eqnarray}
\end{subequations}
Here $\tau_c$ and $\tau_v$ are the momentum relaxation times in the conduction and valence bands, and $p_\omega= \hbar\sqrt{E_{exc}/(\hbar\omega-2\Delta)}/a_B$ with $E_{exc}=\mu e^4/(2\hbar^2\varepsilon^2)$  being the exciton binding energy in three dimensions.

Taking  $d_z \sim 0.1$  e$ \AA$,
$\tau_c \sim\tau_v \sim  10^{-13}$ s  (from the reported mobility $200$ cm$^2/$(V$\cdot$ s)~\cite{Kis}),
and the helicity $\kappa=0.7$, we find the strength of the one monolayer circular PGE signal $\sim 10^{-8} $A$/$W for $\Delta=0.9$ eV, $\hbar\omega=1.95$ eV. This value exceeds the helicity-dependent spin-galvanic signal in 2D GaAs~\cite{Ganichev2001}. The ratio of the net linear PGE and circular PGE is small as $\tau/\tau^{sk}$.

\textit{Side jump photocurrent.} Since the leading ballistic linear PGE, Eq.~(\ref{eq:xi_zeta_bal}b), is inversely proportional to the impurity concentration, in sufficiently high mobility samples it dominates the side jump current. The side jump current, e.g., due to direct optical transitions  to $\zeta$ is obtained using Eqs.~(\ref{eq:delta_v}), (\ref{eq:side_jump}) and the expressions for $V^{K(K^{\prime})}_{-+}(\mathbf{p})$. The result is
\[
\zeta^{dir}_{sj}= 8\left(\frac{e}{\hbar}\right)^2 \frac{Z(p_\omega)\Delta^3}{(\hbar\omega)^3\omega}\,.
\]
Other contributions to $\zeta$ stem from the asymmetry of  impurity-assisted photoabsorption or from the side jumps of photogenerated carriers due to scattering off impurities, and are of the same order of magnitude as $\zeta^{dir}_{sj}$.

\emph{Valley photocurrent.} In addition to the net current, the asymmetric photogeneration leads to the valley currents equal in magnitude but oppositely directed in the $K$ and $K^{\prime}$ valleys, defined by $
    \mathbf{j}^{bal}_v=e\sum_{\mathbf{p},l} (-1)^j \mathbf{v}_{\mathbf{p},l} \delta f_{l}(\mathbf{p})$.  The dominant ballistic contributions to circular and linear valley PGE
can be found using Eqs.~(\ref{eq:collision_balance}), (\ref{eq:photogeneration}) and (\ref{eq:w_skew}):
\begin{equation}
\mathbf{j}^{bal}_v= |\mathbf{E}|^2 \left[ \xi^{bal}\frac{\hbar\omega}{\Delta} \left(\hat{\boldsymbol{z}}\times \hat{S}\mathbf{d}\right) +  \zeta^{bal}\frac{\Delta}{\hbar\omega}\hat{z}\times [\boldsymbol{\kappa}\times\mathbf{d}] \right],
\label{valley_current_value}
\end{equation}
where $\xi^{bal}$ and $\zeta^{bal}$ are given by Eq.~(\ref{eq:xi_zeta_bal}).
The linear and circular valley PGE are related, respectively,  to the net circular ($\xi^{bal}$) and linear ($\zeta^{bal}$) PGE.
Therefore at
$\tau_l/\tau_l^{sk}\sim \delta_l \ll 1$
 the linear valley PGE is the dominant valley current that exceeds the net linear PGE. Valley currents flow perpendicular to the currents (\ref{eq:j_symmetry}),  similar to spin currents in the spin Hall effect.
 Linear valley PGE leads to accumulation of $K$-valley electrons  at the left boundary of the monolayer with respect to the direction of the net linear PGE, and $K^{\prime}$-valley electrons on the right. If intervalley scattering is weak, this accumulation can be measured in transport experiments \cite{condmatYLG}. Valley currents can be possibly also captured experimentally investigating non-local transport \cite{Gorbachev} or non-linear phenomena\cite{Xiao}.

\textit{Magneto-induced photocurrent.}  Magnetic field perpendicular to the layers, $\mathbf{H}=H \hat{\boldsymbol{z}}$, induces a Hall-like current
\begin{equation}\label{eq:magneto-induced_current}
\mathbf{j}^{bal}_H= |\mathbf{E}|^2 \hat{z}\times \left[\xi_H \boldsymbol{\kappa}\times\mathbf{d}+\zeta_H \hat{S}\mathbf{d}\right].
\end{equation}
One obvious contribution to (\ref{eq:magneto-induced_current}) arises from the Lorentz-force term
$\frac{e}{c}\mathbf{v}_l(\mathbf{p})\times\boldsymbol{H}\cdot\frac{\partial\delta f_{l}(\mathbf{p})}{\partial\mathbf{p}}$ included into the left hand side of the Boltzmann equation (\ref{eq:collision_balance}). The corresponding ballistic contributions $\xi^{bal}_H$ and $\zeta^{bal}_H$ to the
 coefficients $\xi_H$ and $\zeta_H$ are related to $\xi^{bal}$ and $\zeta^{bal}$ in Eq.~(\ref{eq:xi_zeta_bal}) by
\begin{equation}
\label{eq:xi_zeta_H}
\xi^{bal}_H= \xi^{bal} \omega_H(\tau_c-\tau_v),
\, \, \zeta^{bal}_H =  \frac{\zeta^{bal} \omega_H\left(\frac{\tau^3_c}{\tau^{sk}_{c}}-\frac{\tau^3_v}{\tau^{sk}_{v}}\right)}{\tau^2_c/\tau^{sk}_{c}+\tau^2_v/\tau^{sk}_{v}},
\end{equation}
where $\omega_H=2eHv^2/\hbar\omega c$ is the cyclotron frequency.

A more interesting mechanism of magneto-induced photocurrent arises from the opposite magnetic field dependence of the band gap in the  $K$ and $K^{\prime}$ valleys;  $\Delta \to \Delta \pm \mathbf{M} \cdot\mathbf{H}$, where $\mathbf{M}$ is the orbital magnetic moment in the Bloch state~\cite{Landau9} at the $K$ or $K'$ points in the Brillouin zone. The latter is related to the Berry curvature~\cite{Landau9} $\mathcal{F}^j_z(\mathbf{p})= \partial_{p_x} \Omega^j_y(\mathbf{p}) - \partial_{p_y} \Omega^j_x(\mathbf{p}) $  and in our system is given by~\cite{niu,Xu}
\begin{equation}\label{eq:magnetic_moment}
    M_z^j= \frac{e\,\mathcal{F}^j_z (\mathbf{p}) \sqrt{\Delta^2+v^2p^2}}{\hbar c}= (-1)^j \frac{ev^2\Delta}{2\hbar c\,(\Delta^2+v^2p^2)}.
\end{equation}
The corresponding contribution to the net ballistic magneto-induced photocurrent may be expressed as
\begin{equation}
\label{mi}
\mathbf{j}_{m}= M_z H \hat{\boldsymbol{z}} \times \frac{\partial \mathbf{j}^{bal}_v}{\partial \Delta},
\end{equation}
where $\mathbf{j}^{bal}_v$ is the magnitude of the $H=0$ ballistic valley current  (\ref{valley_current_value}). 
The magnetic moment contribution  (\ref{mi}) is  $\sim\mathbf{j}^{bal}_v\hbar\omega_H/(\hbar\omega-2\Delta)$, while the Lorentz force contribution to linear PGE (\ref{eq:xi_zeta_H}) is $\sim\mathbf{j}^{bal}_v\omega_H\tau^2_l/\tau_{sk}$. The ratio of $\mathbf{j}_{m}$ to  linear PGE in Eq. (\ref{eq:j_symmetry}) at $H=0$ is
$\hbar\omega_H\tau_{sk}/[(\hbar\omega-2\Delta)\tau]$, which can easily reach $\sim\omega_H\tau$, usually defining the Lorentz force effects.
The role of (\ref{mi}) is further enhanced by the partial cancellation between the Lorentz force contributions of electrons  and holes to linear and circular PGE  in Eq.(\ref{eq:xi_zeta_H}), and the magnetic moment contribution may become the dominant  magneto-induced photocurrent in lower mobility samples.

{\it Discussion}. Besides polar TMD systems, our approach based on Eqs. (\ref{Hamiltonian2D}) and (\ref{light}) may be used to study linear and circular PGE induced by interband transitions in polar boron nitride structures.
 Another interesting system is  a Bernal stacked graphene bilayer placed on a substrate~\cite{McCannFalko}, in which the photocurrents  predicted here can potentially be tuned by gating the system. We note that the existence of  helicity-dependent current induced by an in-plane external magnetic field and  Rashba-like spin-orbit effects \cite{Kane,sog} was recently suggested in graphene  \cite{Sherman}.  We expect that photocurrents in polar bilayer graphene,
  due to the coupling of light to the orbital dipole moment $\mathbf{d}$, will be significantly larger.

\acknowledgments This work was supported by the U. S. Department of Energy Office of Science, Basic Energy Sciences under awards number  DE-SC0010544 (YLG) and
DE-FG02-07ER46452  (S. L. and A. A.). We are grateful to David Cobden, Vladimir Falko, Boris Spivak and Xiaodong Xu for useful discussions.

\end{document}